\documentclass[a4paper]{jpconf}
\usepackage{graphicx}
\usepackage{bm}
\usepackage{cite}
\begin{document}
\title{Specific-heat study for ferromagnetic and antiferromagnetic phases in SrRu$_{\bm{1-x}}$Mn$_{\bm x}$O$_{\bm 3}$}
\author{A Yamaji$^{1,\dag}$, M Yokoyama$^{1,\ddag}$, Y Nishihara$^1$, Y Narumi$^2$ and K Kindo$^3$}
\address{$^1$Faculty of Science, Ibaraki University, Mito 310-8512, Japan}
\address{$^2$Institute for Materials Research, Tohoku University, Sendai 980-8577, Japan}
\address{$^3$Institute for Solid State Physics, The University of Tokyo, Kashiwa 270-8581, Japan}
\ead{$^\dag$08nm234x@mcs.ibaraki.ac.jp, $^\ddag$makotti@mx.ibaraki.ac.jp}

\begin{abstract}
Low-temperature electronic states in SrRu$_{1-x}$Mn$_x$O$_3$ for $x \le 0.6$ have been investigated by means of specific-heat $C_p$ measurements. We have found that a jump anomaly observed in $C_p$ at the ferromagnetic (FM) transition temperature for SrRuO$_3$ changes into a broad peak by only 5\% substitution of Mn for Ru. With further doping Mn, the low-temperature electronic specific-heat coefficient $\gamma$ is markedly reduced from the value at $x=0$ (33 mJ/K$^2$ mol), in connection with the suppression of the FM phase as well as the enhancement of the resistivity. For $x\ge 0.4$, $\gamma$ approaches to $\sim 5$ mJ/K$^2$ mol or less, where the antiferromagnetic order with an insulating feature in resistivity is generated. We suggest from these results that both disorder and reconstruction of the electronic states induced by doping Mn are coupled with the magnetic ground states and transport properties.
\end{abstract} 

\section{Introduction}
The interplay of magnetism and charge transport in the vicinity of metal-insulator transition is one of the most intriguing issues in the physics of strongly correlated electron systems. The distorted perovskite compound SrRuO$_3$ shows a ferromagnetic (FM) order below $T_C=160\ {\rm K}$, in which the itinerant 4d electrons are considered to be responsible for the spontaneous spin polarization \cite{rf:Callaghan66,rf:Kanbayasi76,rf:Allen96}. The electrical resistivity $\rho$ exhibits unusual metallic behavior called ``bad metal", characterized by an absence of suppression in $\rho$ and a very small mean-free pass comparable to lattice constants at high temperatures \cite{rf:Allen96}. In addition, optical conductivity \cite{rf:Kostic98} and photoemission \cite{rf:Okamoto99} studies indicate the enhancements of anomalous electronic states originating from many-body correlation effects. 
%These results strongly suggest a close relationship between the electron conductions and the magnetism.  

In the mixed compounds SrRu$_{1-x}$Mn$_x$O$_3$ \cite{rf:Sahu2002,rf:Cao2005,rf:Yokoyama2005,rf:Han2006,rf:Zhang2006,rf:Woodward2008,rf:Kolesnik2008}, it is revealed that the substitution of Mn for Ru suppresses the FM phase, and then induces the C-type antiferromagnetic (AFM) phase above $x_c\sim 0.3-0.4$. The structural transition from orthorombic to tetragonal symmetries also occurs in connection with the variation of magnetic ground state. Furthermore, the Mn substitution changes the characteristic of $\rho$ from metallic to insulating ones. These features suggest that doping Mn into SrRuO$_3$ modifies the itinerant electronic states due to the effects of strong correlations between Ru 4d and Mn 3d electrons, and it significantly affects the magnetic, transport and lattice properties. It is therefore interesting to investigate the relationship between the electronic states and these properties. To clarify this, we have performed specific-heat measurements on SrRu$_{1-x}$Mn$_x$O$_3$ in the low and intermediate $x$ ranges.

\section{Experimental Procedure}
Polycrystalline samples of SrRu$_{1-x}$Mn$_x$O$_3$ with $0 \le x \le 0.6$ were prepared by means of the conventional solid-state method. The mixtures of appropriate amounts of SrCO$_3$, RuO$_2$ and MnO are first calcined at 750 $^\circ$C for 4 hours. They were shaped into pellets after careful mixing, and then sintered at 1300 $^\circ$C for 24 hours. This sinter process was iterated 10 times to achieve homogeneity of the samples. The details on the sample preparation are presented elsewhere \cite{rf:Yokoyama2005}. Specific-heat $C_p$ was measured between 2 K and 275 K with a thermal-relaxation technique. To check an effect of thermal conductance in the sintered samples on the $C_p$ data, we performed the experiments using both a commercial system (PPMS: Quantum Design) and a hand-made equipment, where we used the plate-shaped samples with the mass of about 4 mg and 80 mg, respectively. The thermal-relaxation curves for all the measurements were well definitive, and the $C_p$ data obtained from both the equipments were consistent within the experimental accuracy. Electrical resistivity measurements were performed using a standard four-wire dc technique from 3.5 K to 300 K. Ac-susceptibility was measured in the temperature range of $5-300\ {\rm K}$ for $0 \le x \le 0.3$ to estimate the FM transition temperature $T_C$. Frequency and amplitude of the applied ac field were 180 Hz and $\sim 0.5\ {\rm Oe}$. 

\section{Results and Discussion}
\begin{figure}[tbp]
\begin{center}
\includegraphics[keepaspectratio,width=0.8\textwidth]{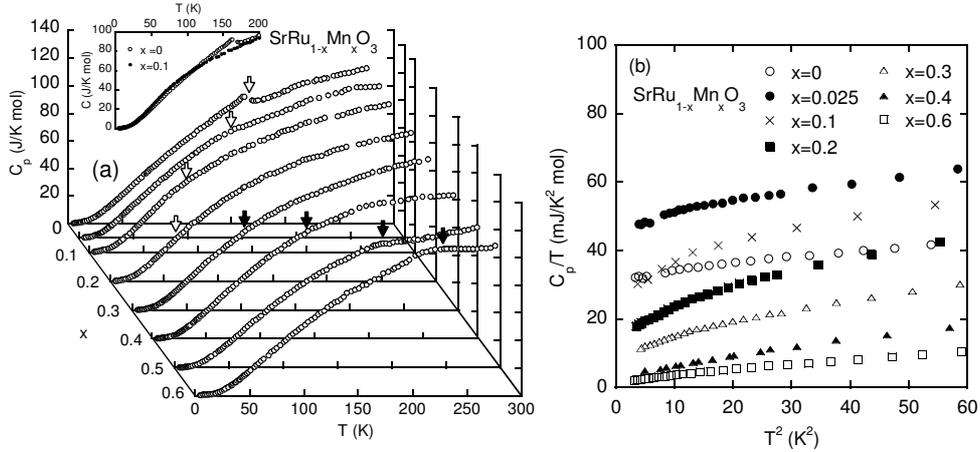}
\end{center}
  \caption{
(a) Temperature variations of specific heat $C_p(T)$, and (b) specific heat divided by temperature versus $T^2$ for SrRu$_{1-x}$Mn$_x$O$_3$ with $0 \le x \le 0.6$. In (a), the open arrows indicate the FM transition temperatures estimated from ac-susceptibility measurements, and the closed arrows show the onsets of AFM phase obtained by the neutron scattering and dc-magnetization experiments \cite{rf:Yokoyama2005}. The comparison of the $C_p(T)$ curves for $x=0$ and 0.1 are shown in the inset of (a). 
}
\end{figure}
Figure 1(a) shows the temperature variations of specific heat $C_p(T)$ for SrRu$_{1-x}$Mn$_x$O$_3$ with $x\le 0.6$. A clear jump associated with the FM transition is observed at 162 K for pure SrRuO$_3$. It is highly suppressed by doping Mn, but the enhancements with very small magnitude are still observed in the wide temperature ranges around $130\ {\rm K}$ and $80\ {\rm K}$ in the $C_p(T)$ curves for $x=0.05$ and $x=0.1$, respectively (see inset of Fig.\ 1). Since these temperatures agree with $T_C$ estimated from temperature variations of ac-susceptibility, the enhancements seen in $C_p(T)$ are considered to be attributed to the FM transition.  On the other hand, $C_p(T)$ shows no significant anomaly at the transition temperature for $0.2 \le x \le 0.4$, which may be relevant to the suppression of the ordered moments in the vicinity of $x_c$ \cite{rf:Yokoyama2005,rf:Kolesnik2008}. Shoulder-like anomalies associated with the AFM transition occur in $C_p$ at $\sim 190\ {\rm K}$ ($x=0.5$) and $\sim 220\ {\rm K}$ ($x=0.6$). The anomaly becomes more pronounced with increasing $x$, due to the development of the AF staggered moment.

In Fig.\ 1(b), we show the specific heat divided by temperature $C_p/T$ at low temperatures, plotted as a function of $T^2$. $C_p/T$ for pure SrRuO$_3$ is roughly proportional to $T^2$ with a finite value for $T\to 0$, indicating that the quasi-particles with the Fermi-liquid excitations as well as phonons are mainly responsible for $C_p$. $C_p/T$ is enhanced by only 2.5\% substitution of Mn for Ru, and it shows a slight deviation from the $T^2$ dependence below $\sim 3.6\ {\rm K}$ ($T^2\sim 13\ {\rm K}^2$). The magnitude of $C_p/T$ decreases with further doping Mn. For $0.1 \le x \le 0.25$, the downward deviations from the $T^2$ function are clearly observed in the $C_p/T$ curves, strongly suggesting that the component proportional to $T^n$ ($1 < n < 3$) is involved in $C_p(T)$ at low temperatures. In the AFM region, $x\ge 0.3$, the $T^n$ component in $C_p/T$ is reduced with increasing $x$, and the $T^2$ dependence is again dominant above $x=0.4$.    

\begin{figure}[tbp]
\begin{center}
\includegraphics[keepaspectratio,width=0.4\textwidth]{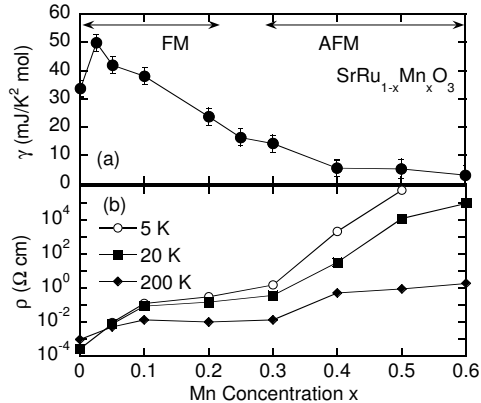}
\end{center}
  \caption{
(a) Low-temperature electronic specific-heat coefficient $\gamma$ and (b) electrical resistivity $\rho$ at 5 K, 20 K, and 300 K for SrRu$_{1-x}$Mn$_x$O$_3$ ($0 \le x\ \le0.6$). Note that vertical axis in (b) is indicated with the logarithmic scale.
}
\end{figure}
The $C_p/T$ data are fitted with a function of $\gamma +\beta T^2$ to derive the electronic specific-heat coefficient $\gamma$. Displayed in Fig.\ 2(a) are the $x$ variations of $\gamma$ obtained from the best fit in the temperature range of $5-8\ {\rm K}$. We attempted to plot the $\gamma$ values obtained by fitting in the lower temperature range as well as the $C_p/T$ data at 2 K in order to check the effect of the $T^n$ component on the present analysis, and confirmed that these quantities exhibit qualitatively the same $x$ dependence as that seen in Fig.\ 2(a). For comparison, the $x$ variations of $\rho$ at 5 K, 20 K and 200 K are also plotted in Fig.\ 2(b). The $\gamma$ value for pure SrRuO$_3$ is estimated to be 33 mJ/K$^2$ mol, which is roughly consistent with the previously reported ones \cite{rf:Allen96,rf:Okamoto99,rf:Cao97}. In SrRu$_{1-x}$Mn$_x$O$_3$, we have found three features in the $x$ variations of $\gamma$ and $\rho$. First, $\gamma$ for $x \le 0.1$ is weakly enhanced and shows a maximum at $x \sim 0.025$. In this $x$ range the difference in resistivity $\Delta\rho=\rho(5\ {\rm K})-\rho(200\ {\rm K})$ changes from negative to positive values through the zero at $x\sim 0.05$. Secondary, $\gamma$ linearly decreases with increasing $x$ up to $\sim 0.4$, accompanied by a gradual increase in $\rho$ at low temperatures. Third, $\gamma$ for $0.4 \le x \le 0.6$ is suppressed to $\sim 5$ mJ/K$^2$ mol or less, and shows a tendency to be independent of $x$, where the insulating behavior with the strongly enhanced resistivity is observed at 5 K. On the other hand, the Debye temperatures estimated from the $\beta$ parameters were $\sim 350\ {\rm K}$ for all the $x$ range presently investigated. 

It is found that the jump anomaly at $T_C$ observed in $C_p$ for $x=0$ is rapidly suppressed and changes into the broad peak for $x\le 0.1$. At the same time, both $\Delta\rho$ and $\rho$ increase, while $\gamma$ stays in large values. Since the dc magnetization \cite{rf:Cao2005,rf:Kolesnik2008} and ac susceptibility measurements indicate that the FM phase is still stable in this $x$ range, these variations are suggested to be attributed to the breakdown of coherency for the itinerant Ru 4d electrons generated locally at the Mn ions, that is caused by the difference of spin and orbital states and ionic radius between Ru$^{4+}$ and Mn$^{4+}$ \cite{rf:Sahu2002,rf:Cao2005,rf:Yokoyama2005,rf:Han2006,rf:Zhang2006,rf:Woodward2008,rf:Kolesnik2008}. As $x$ is increased up to $\sim 0.3$, $\rho$ gradually grows in cooperation with a reduction of $\gamma$. In addition, the evolution of the $T^n$ component is observed in $C_p(T)$ at low temperatures. These features in $\gamma$, $\rho$ and  $C_p(T)$ may be coupled with the magnetic fluctuations enhanced through the competition between FM and AFM orders. In particular, the $T^{3/2}$ dependence in $C_p$ is expected to occur in the usual three-dimensional ferromagnet with the localized spins. On the other hand, $\rho$ at 5 K becomes extremely large and the magnitude of $\gamma$ is highly reduced between $x\sim 0.4$ and 0.6. Such large variations in $\rho$ and $\gamma$ indicate that the electronic state is reconstructed due to the strong hybridization and correlation between the Ru 4d and Mn 3d electrons in this $x$ range, which plays a crucial role in the evolutions of the insulating state, the long-range AFM order, and the tetragonal crystal structure. To clarify the magnetic correlations in the FM and AFM states, we plan to perform the neutron scattering experiments in the entire $x$ range.  

\section{Summary}
We performed $C_p$ measurements for SrRu$_{1-x}$Mn$_x$O$_3$ with $0 \le x \le 0.6$. It is observed that doping Mn into SrRuO$_3$ obscures the jump in $C_p$ associated with the FM transition, and yields unusual development of $\gamma$ for $x \le 0.1$. For $x\ge 0.1$ $\gamma$ linearly decreases with increasing $x$. Above $x \sim 0.4$, the magnitude of $\gamma$ becomes very small, and the shoulder-like anomalies develop in $C_p$ at the onsets of the AFM transition. We discuss the relationship between $\gamma$ and $\rho$ in the small and intermediate $x$ ranges, and suggest that the magnetic and transport properties are affected by the changes in the electronic states induced by both the disorder and correlation effects.

\ack
We thank T. Koseki, H. Hagiya and H. Kawanaka for helpful discussion. This work was carried out by the joint research in the Institute for Solid State Physics, the University of Tokyo.

\end{document}